\documentclass[prc,twocolumn,showpacs,preprintnumbers,amsmath,amssymb]{revtex4-1}
\usepackage[utf8]{inputenc}
\usepackage{graphicx}
\usepackage{tabularx}

\usepackage{bm}
\usepackage{array}
\usepackage{xcolor}

\usepackage{color}

\usepackage[normalem]{ulem}  
\renewcommand\sout{\bgroup \color{red} \ULdepth=-.5ex \ULset}
\usepackage{hyperref}

\allowdisplaybreaks

\begin{document}

\title{Ridge regression for minimizing hyperon resonances' couplings in the $K^+ \Lambda$ photoproduction}

\author{D.~Petrellis, D.~Skoupil}

\affiliation{Nuclear Physics Institute, CAS, \v{R}e\v{z}/Prague, Czech Republic}

\date{\today}

\begin{abstract}
We employed the isobar model for investigating the $K^+\Lambda$ photoproduction process. We paid special attention to the recent CLAS polarization data and enhanced the $\chi^2$ minimization by adding a penalty term. Without changing the set of included resonances used by the model, this technique known as Ridge regression leads to reduced couplings that in previous studies acquired unreasonably large values. As a result, we have arrived at a much more robust model with hyperon couplings which are reduced to more physical values. This model serves us to extract valuable information on the background to the $K^+\Lambda$ photoproduction and particularly on the role of various hyperon resonances. The set of the nucleon resonances is the same with respect to previous fits but their role may have changed due to different couplings which they acquire in the present fit.
\end{abstract}

\maketitle

\section{Introduction}

The study of photoproduction and electroproduction of hyperons from nucleons in the third nucleon resonance region gives us important information about the spectrum of baryon resonances and interactions in systems of hyperons and nucleons which arise from QCD. Besides learning more about the reaction mechanism, a correct description of the elementary production process is important for getting reliable predictions of the excitation spectra for production of $\Lambda$ hypernuclei~\cite{hyper}.


The nucleon resonances and their properties were studied quite thoroughly by using the kaon photoproduction process. Over the past few decades, a number of theoretical studies of the hyperon production have been performed and they focused mainly on the process of $K^+\Lambda$ photoproduction. Let us here briefly present the body of research on $K^+\Lambda$ photoproduction done in the past. First studies of this topic were performed in 1960s when Kuo~\cite{Kuo} and then Thom~\cite{Thom} published their phenomenological analyses of the $p(\gamma,K^+)\Lambda$ reaction. When more data had become available, several new studies of the $K^+\Lambda$ process were performed in the 1980s~\cite{ABW, Rosenthal} and also in the 1990s~\cite{WJC,AS,SL,Mizutani} and the studies focused also on the problem of missing nucleon resonances~\cite{Mart-Bennhold}. At the end of the century, the study of meson photoproduction at energies above 4~GeV became experimentally accessible which motivated development of a model for high energy kaon photoproduction~\cite{Guidal}. After the turn of the century, the theory group at the Ghent University, studied the effects of background contributions to the $K^+\Lambda$ photoproduction~\cite{GentIM} and attempted at extracting the information on nucleon resonances from the then still somewhat limited data set~\cite{Janssen-2003}. Most importantly for the present paper, they also examined the role of hyperon resonances in the $K^+\Lambda$ photoproduction~\cite{Janssen-2001EPJ}. After the publication of the CLAS data on differential cross sections and a few polarization observables~\cite{CLAS05}, new fits to these data were performed~\cite{Puente}. In the following years, the aforementioned Ghent group developed a model capable of simultaneously describing the process near the threshold and also at high energies~\cite{Corthals,DeCruz}.  Worth mentioning are also the coupled-channel analyses~\cite{Shklyar, Shyam, Julia, Borasoy}, particularly the extensive work by the Bonn-Gatchina~\cite{BG1,BG2,BG2012,BG3}, Juelich-Bonn~\cite{Roenchen}, and ANL-Osaka~\cite{Kamano} groups. Even though the body of research done in the past is quite substantial, extracting resonances in baryon spectroscopy still remains a very demanding task.

There are number of studies focusing on the spectrum of nucleon resonances  in the $K^+\Lambda$ photoproduction while the amount of studies on hyperon resonances is somewhat limited~\cite{Mart-are}. In an ideal case, this kind of study should be done in a process where the hyperon resonances propagate in the $s$ channel, such as kaon-nucleon scattering. An experiment on kaon-nucleon scattering became one of the main subjects to be studied at JPARC, Japan~\cite{JPARC}. However, one needs a very intense kaon beam in order to be able to achieve accuracy which was obtained in experiments with electromagnetic beams. Moreover, it is quite well-known that the hadronic interaction is not as easy to grasp as the electromagnetic interaction, which further impedes extracting information on hyperon resonances from this process. With this in mind, one can say that the photo-induced process of kaon production could become a valuable source of knowledge on hyperon resonances. They are being exchanged in the $u$ channel of the process and thus their contributions are sensitive to the observables at backward kaon angles.

Models based on effective Lagrangians belong to the most effective tools we have to describe the $K^+\Lambda$ photoproduction. Since there is no explicit connection to QCD in these models, the number of parameters is related to the number of resonances included. In the case of kaon photoproduction this number is relatively large. In the isobar-model description of the kaon photoproduction process, there has been a long-standing issue of having too large coupling constants of the hyperon resonances. In many isobar models, the hyperon resonances' couplings acquired values larger than one, for example the Saclay-Lyon model~\cite{SL}, the model by Williams, Ji, and Cotanch~\cite{WJC} and the model of Janssen \emph{et al.}~\cite{GentIM} introduce spin-1/2 hyperon resonances with couplings larger than one. In 2016, we analysed the data, available at the time, for $K^+\Lambda$ photoproduction which resulted in two new models called BS1 and BS2~\cite{BS2016} and also in these models the hyperon couplings tend to be large. At the tree-level approximation, which we use, couplings larger than one are still acceptable but they are not reasonable in view of the philosophy of perturbation calculations. Furthermore, it can be said that the model assumptions with respect to the way the background terms are treated influence the extracted information about the resonant terms.

A year after the publication of BS1 and BS2 models, the $K^+\Lambda$ data base was replenished by CLAS data~\cite{Paterson} on photon-beam asymmetry $\Sigma$, target asymmetry $T$, and beam-recoil polarization asymmetries with linearly polarized photons, $O_{x^\prime}$ and $O_{z^\prime}$. It was these polarization data that motivated us to look once again at the $K^+\Lambda$ channel trying to disentangle the spectrum of hyperon resonances contributing to this process.

Another reason for analyzing the $K^+\Lambda$ channel again was a more thorough fitting method which penalizes the introduced parameters for their values. This technique, known as Ridge regularization, helps us to solve the issue of unphysically large hyperon couplings. Regularization techniques are commonly used in statistics and machine learning in order to prevent overfitting and produce models that generalize better to new data. Although this is usually achieved by making the model sparser, Ridge regression allows one to still work within a certain model -- a given set of resonances, in our case -- and mitigate the values of its fitted parameters. We demonstrate this method by applying it on two models which were derived from the BS2 model.

This paper is organised as follows: In Sec.~\ref{sec:methodology}, we show the method we use in this study for describing the $K^+\Lambda$ photoproduction off the proton. In Sec.~\ref{sec:fitting}, we introduce the reader to the method which we use in order to adjust the values of model parameters to experimental data. In Sec.~\ref{sec:results}, we discuss our results and in Sec.~\ref{sec:conclusion} we give a short summary and conclusion.



\section{Formalism of the isobar model}
\label{sec:methodology}

In this study, we use an isobar model where the amplitude is constructed from effective meson-baryon Lagrangians. The nonresonant part of the amplitude consists of exchanges of the ground-state hadrons (Born terms) and exchanges of resonances in the $t$ ($K^*$ and $K_1$) and $u$ channel ($\Lambda$ and $\Sigma$ resonances), so-called non Born terms. The resonant part of the model is given by $s$-channel exchanges of nucleon resonances with masses from around the threshold of the process to approximately 2.2~GeV. The contributions beyond this tree-level order, such as rescattering or interactions in the final state, are neglected in this approach. 

Since the exchanged particles are not point-like, we introduce a hadronic form factor in the strong vertex. Besides accounting for the extended structure of exchanged particles, the role of the hadronic form factors is to regularise the amplitude at large energies. Too large a contribution of the Born terms to the cross sections is one of the characteristic features of the isobar-model description of the $p(\gamma,K^+)\Lambda$ process. Introducing hadronic form factors in the strong vertices is one way to deal with this issue. The other way of suppressing Born terms is to introduce exchanges of hyperon resonances in the $u$ channel. These resonances, which we pay special attention to in this paper, can also play an important role in the dynamics, as shown e.g. in the Saclay-Lyon model~\cite{SL}. What is more, the presence of hyperon resonances can substantially improve the agreement of the model prediction with data, reduce the $\chi^2$ value, and shift the value of the hadron cut-off parameter to a harder region.

The process of photoproduction of kaons occurs in the so-called third nucleon resonance region, where there are plenty of excited states of the nucleon and none of these resonances dominates. Therefore, we have to take into account \emph{a priori} more than 20 nucleon, kaon, and hyperon resonances. This leads to a large number of resonance combinations that describe the data in an acceptable way with a reasonably small $\chi^2$. In order to reduce the amount of acceptable models, we impose constraints on the values of the coupling constants in the $K\Lambda N$ and $K\Sigma N$ vertices, relating them to the well-known $\pi NN$ value by means of the SU(3) symmetry~\cite{BS2016}.


One of the most important ingredients of our model is the consistent formalism for the exchange of high-spin resonances. The Rarita-Schwinger description of high-spin fermion fields includes nonphysical degrees of freedom which are connected to their lower-spin content. If the Rarita-Schwinger field is off its mass shell, the nonphysical parts can, in principle, participate in the interaction, which is the reason to call it "inconsistent". In our work, we use interaction Lagrangians which are invariant under the local U(1) gauge transformation of the Rarita-Schwinger field~\cite{Pascalutsa,Vrancx}. This property then removes all nonphysical contributions of lower-spin components from the amplitude.

Since in the isobar model we construct the amplitude with tree-level Feynman diagrams only, the unitarity is broken. In order to restore it, we use the energy-dependent decay widths of the nucleon resonances. The energy dependence of the width is given by the possibility of a resonance to decay into various open channels. 

As the purpose of the current paper is not to reintroduce our model, see Refs.~\cite{BS2016} and~\cite{BS2018} for more details.



\section{Fitting procedure}
\label{sec:fitting}

In both BS1 and BS2 models there is a significant overlap in the set of nucleon resonances but also a substantial difference in the set of hyperon resonances (for details, see Tab.~II in ~\cite{BS2016}), which leads to a different description of background.


In the BS1 model, there is only one hyperon resonance [$\Lambda(1520)3/2^-$] with its couplings below one (in the absolute value), the rest of the hyperon couplings rise well above one. Similarly, in the BS2 model there is only one hyperon resonance [$\Sigma(1940)3/2^-$] whose couplings are below one, the remaining hyperon resonance's couplings are above one. From what we see, hyperon resonances whose couplings do not rise above one are spin-3/2 resonances. It is typical for couplings of spin-1/2 hyperon resonances to acquire values well above one.

Up to now, the general approach in dealing with the high values of the hyperon couplings, that was also followed in~\cite{BS2016}, has been to impose constrains on their values during the error minimization procedure. The regularization approach that we propose in the current paper provides a less arbitrary solution to this problem.

\subsection{Ridge regularization}
\label{sub:Ridge}
As already mentioned, regularization is the standard approach used in machine learning in order to prevent a model from overfitting the data ~\cite{Bishop, Hastie}. This is achieved by adding to the error function a penalty term that contains powers of the absolute values of the parameters. The regularized error function thus becomes

\begin{equation}
  \chi^2_{P} = \chi^2 + P(\lambda),
  \label{eq:chi^2_T}
\end{equation}
with
\begin{equation}
  \chi^2 = \sum_{i=1}^{N}\frac{[d_i-p_i(w_1,\ldots,w_m)]^2}{(\sigma_{d_i})^2},
  \label{eq:chi^2}
\end{equation}
the ordinary $\chi^{2}$ error function, where $d_i$ represents a data point, $p_i$ the corresponding prediction of the model, depending on a set of parameters $\{w_1\dots w_m\}$ and $\sigma_{d_i}$ the error connected with each measurement. The penalty term
\begin{equation}
  P(\lambda) = \lambda \sum_{j=1}^{m}|w_j|^{q},
  \label{eq:Pen}
\end{equation}
contains an $L_q$ norm of the parameter vector and, in effect, converts the problem of ordinary error minimization to one of constrained minimization, where the parameters are not allowed to take arbitrarily large (absolute) values that would cause the model to fit the noise in the sample.

The regularization parameter $\lambda$ determines the magnitude of suppression of the parameter values, while the power $q$ determines its character. In the two most commonly used cases $q=1$ and $q=2$ -- known in the literature as LASSO (least absolute shrinkage and selection operator) and Ridge regression, respectively -- the constraint affects the position of the minimum differently.

\begin{figure}
    \centering
    \includegraphics[width=0.45\textwidth]{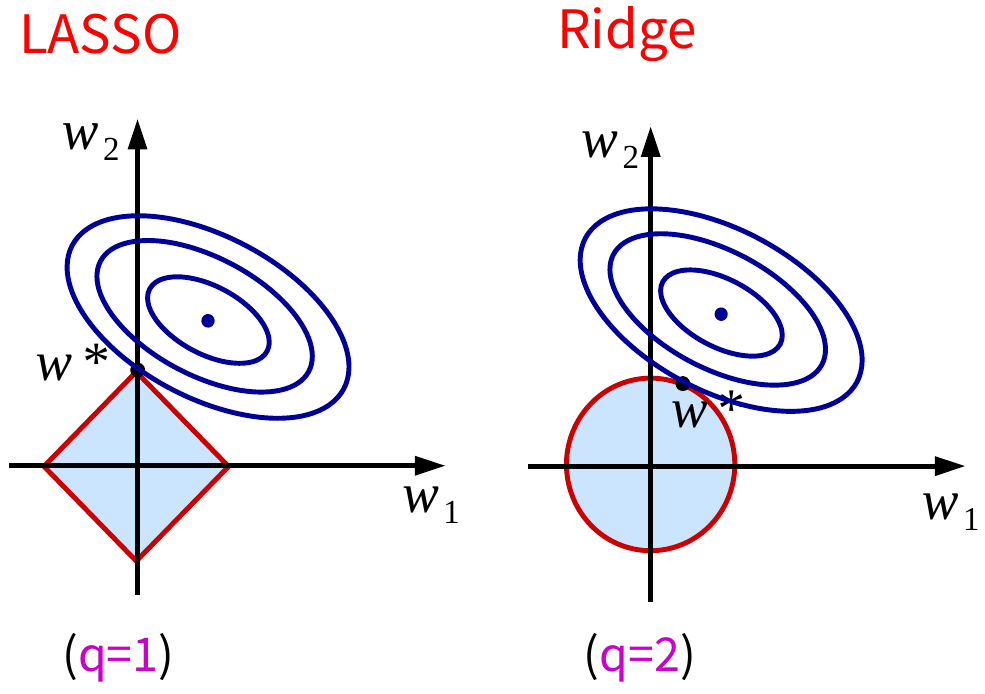}
    \caption{Schematic illustration of the effect of LASSO and Ridge regularization on the optimal parameter values. The ellipses represent the contours of the error function with the \textit{free} minimum in the middle, while the colored area around the origin represents the constraint.  The point $w^{*} = (w^{*}_{1},w^{*}_{2})$ is the \textit{constrained} minimum, and it can be seen that in the case of LASSO $w^{*}_{1} = 0$, while with Ridge $w^{*}_{1}\approx 0$.}
    \label{fig:Ridge}
\end{figure}

As can be seen in Fig.~\ref{fig:Ridge}, with LASSO, due to the geometry of the constraint, some parameters are forced to take zero values, while with Ridge they approach, but do not become exactly zero. This tendency  of LASSO to drive some parameters to zero, thus favoring sparser models, makes it a suitable tool for model selection~\cite{Guegan, Landay1, Landay2, Bydzovsky}. 

Ridge, on the other hand, shrinks the values of the parameters, without annihilating them, so their number is not affected. The penalty term can also have $1<q<2$ and thus combine features of both LASSO and Ridge, without the problems that arise from the points of non-differentiability encountered in LASSO ~\cite{Hastie}. One reason Ridge is more relevant in our case than LASSO, is that the set of resonances included in the BS2 model have been chosen according to a previous model selection study which had been motivated by a robust Bayesian analysis~\cite{DeCruzPhD}. The second model that we use to demonstrate this technique is a variant of BS2 based on results from Ref.~\cite{Mart-are}.

Therefore, the penalty term that we used in our study is specifically written as 
\begin{equation}
  P(\lambda) = \lambda^{4} \sum_{j=1}^{m}g_j^{2},
  \label{eq:Pen2}
\end{equation}
where $g_j$ are the couplings of the resonances included in our model and $m$ is the number of couplings. Please note that the number of parameters exceeds the number of resonances since a spin-1/2 resonance introduces one free parameter whereas a spin-3/2 resonance introduces two free parameters.

\begin{figure}
    \centering
    \includegraphics[width=0.45\textwidth]{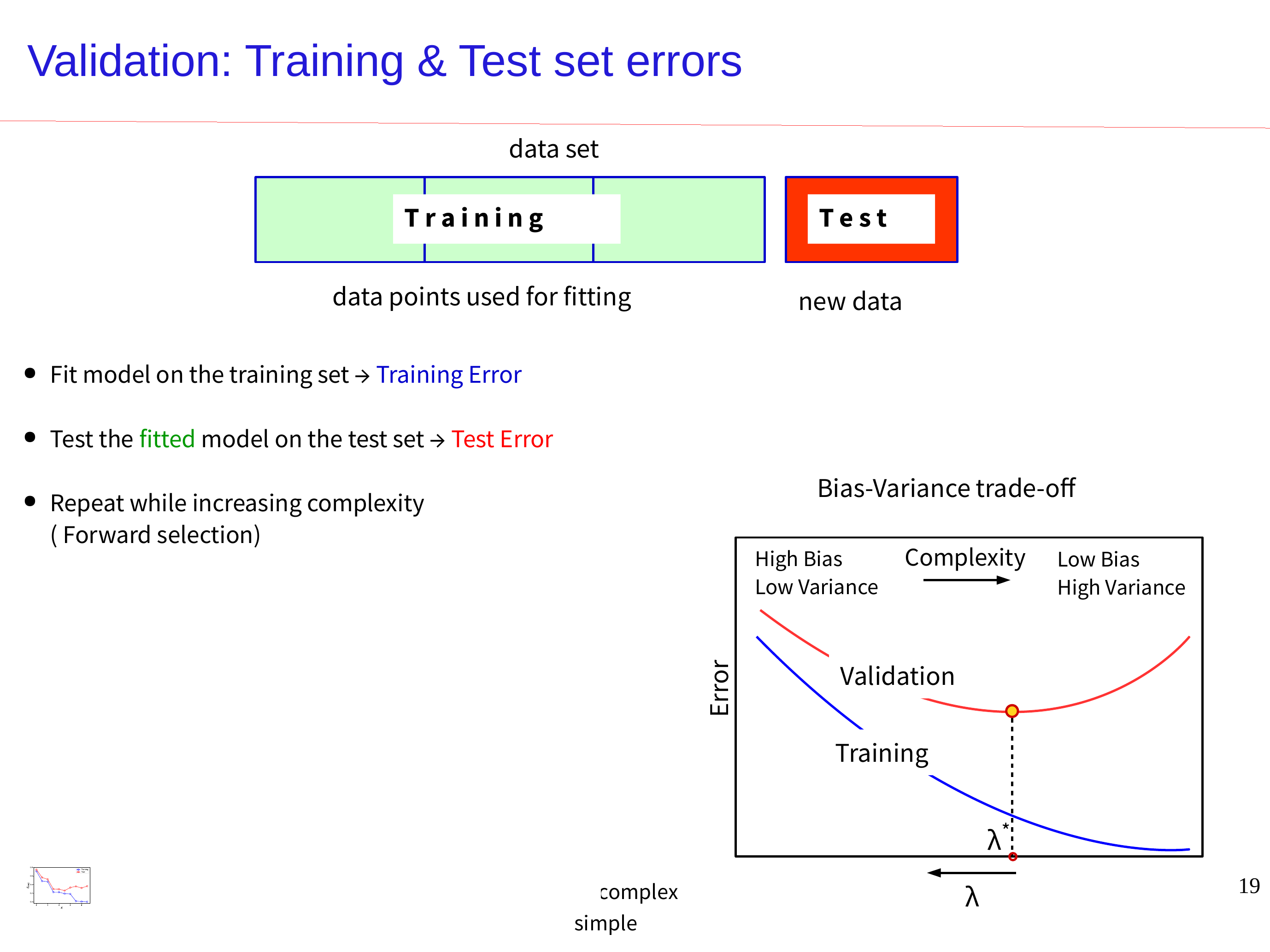}
    \caption{Schematic illustration of the evolution of the training and validation errors with increasing model complexity. Please note that the regularization parameter $\lambda$ increases from right to left.}
    \label{fig:Bias_Var}
\end{figure}

The presence of the regularization term introduces some bias in the model and will increase the error anyway; however, if $\lambda$ is too large the penalty on the parameters will be too high, causing the model to underfit the data, while if it is too small it will fail to prevent overfitting the data. Apparently, the optimal value of $\lambda$ must represent a balance between these two extremes. This is done by examining a set of $\lambda$ values from an appropriately chosen range and applying cross validation for each value, as we will see in the next section. Using $\lambda^{4}$ as a regularization parameter and varying $\lambda$ in equidistant steps, for values smaller than one, allows us to focus more on the region of small $\lambda$, which is more relevant in our case.  

\subsection{Cross validation}
\label{sub:cv}
Cross validation is a model assessment tool, whereby the performance of a model is evaluated according to how well it describes new data, i.e. data that have not been used for fitting. To that end, the data set is split randomly into a training set used to fit the model and a validation set used only to test the quality of the fitting. The model with the lowest validation error is the preferred one. 

Even though Ridge does not produce a new model by changing the number of parameters like LASSO, it clearly affects the model's complexity by restraining its parameters and the optimal $\lambda$ can be chosen as the one that yields the lowest error on the validation set. More precisely, we start with a $\lambda_{\textrm{max}}$ and reduce it in steps, where in each step the model is fitted on the training set and the resulting parameter values are used to calculate the error on the validation set. Those parameter values are passed as starting values in the next step where the training set is fitted with the new value of $\lambda$ and this process is repeated until a $\lambda_{\textrm{min}}$ (usually $\lambda_{\textrm{min}}=0$) is reached.

As shown schematically in Fig.~\ref{fig:Bias_Var} the training error decreases with decreasing $\lambda$, as the constraint on the parameters is relaxed, while the validation (or prediction) error, which is consistently higher, decreases up to a certain $\lambda$ value, below which it starts increasing. That is the point where the model starts overfitting the training set and as a result is getting worse at predicting the validation set. Thus, the minimum of the validation error indicates the optimum value of $\lambda$.

Due to the arbitrariness in the division into training and validation set, the technique known as $k$-fold cross validation is employed. The original data set is divided into $k$ parts of equal size and cross validation is performed $k$ times, using each of the $k$ parts for validation and the rest of the data for training, every time. 

Since each of these cross-validation runs yields a different outcome, we end-up with $k$ different results for the training error and $k$ different results for the validation error, \textit{for each value of $\lambda$}. The averages of these values provide estimates of the training and validation errors corresponding to the  given $\lambda$. In particular, we are interested in the errors on the validation set since their minimum will help us determine the optimal value of $\lambda$, as demonstrated in Fig.~\ref{fig:Bias_Var}. If we denote the validation error corresponding to the $l$th run by $CV_l(\lambda)$, then the average over all $k$ runs, for a certain $\lambda$, can be written as
\begin{equation}
\overline{CV}(\lambda) = \frac{1}{k}\sum_{l=1}^{k}CV_l(\lambda)
  \label{eq:CV}
\end{equation}
and the optimal $\lambda$, as discussed above, is

\begin{equation}
  \lambda^{*} =  \underset{\lambda \in \lbrace  \lambda_{\textrm{min}},...,\lambda_{\textrm{max}} \rbrace }{\arg\!\min} \overline{CV}(\lambda) 
  \label{eq:lmin}
\end{equation}

In order to take into account the uncertainty associated with the estimation of the validation error and subsequently of the optimal value of $\lambda$, cross validation is often accompanied by the ``one standard-error" rule. According to this rule (see ~\cite{Hastie} and ~\cite{Landay1}), the most parsimonious model within one standard error from the minimum of the validation error should be chosen. This implies, in our case, an optimal $\lambda$ that corresponds to one standard-error above the minimum. 

\begin{figure}[h]
\centering
\includegraphics[width=0.45\textwidth]{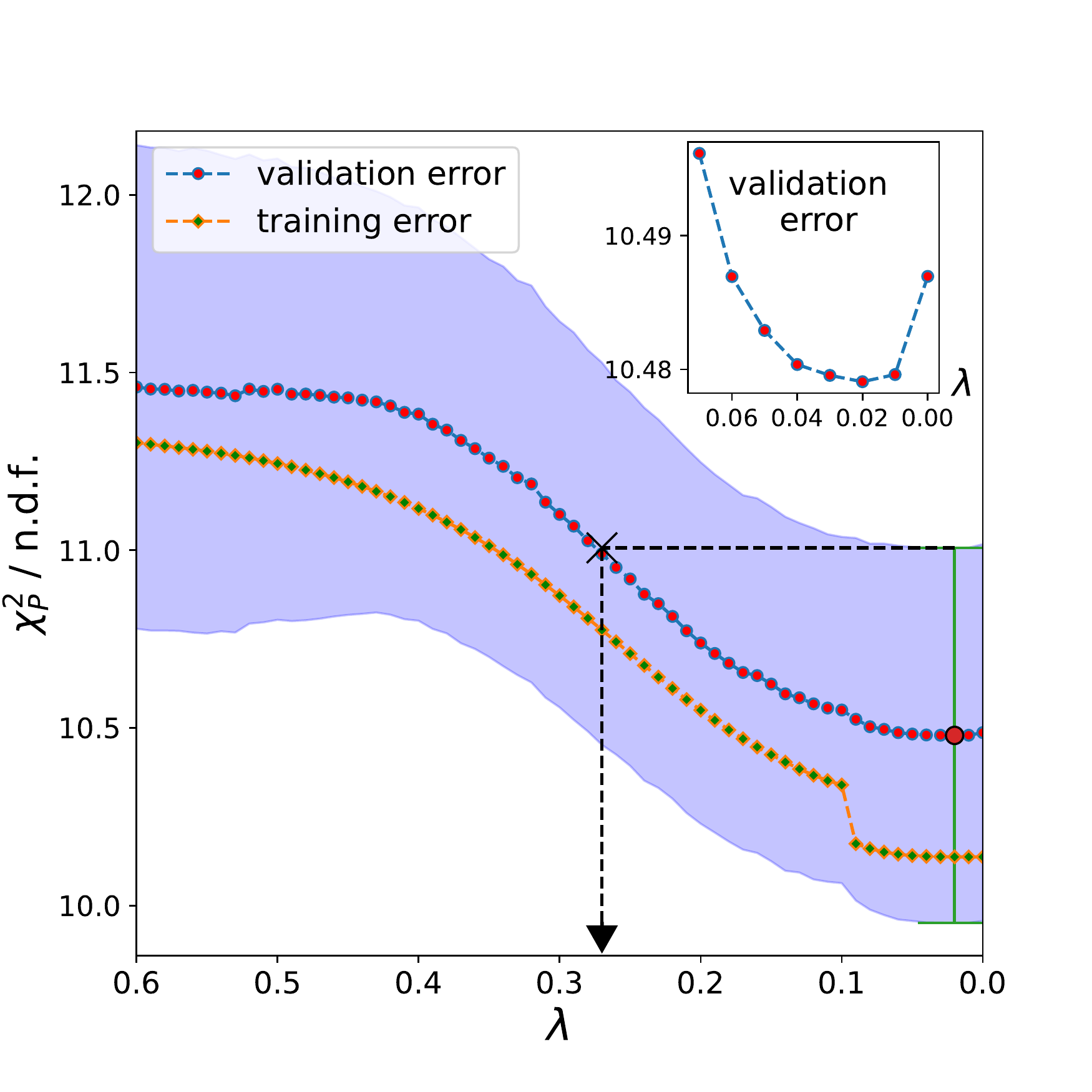}
\caption{Results of 3-fold cross validation to determine the optimal $\lambda$ value used in the BS2r fit. The inset shows the position of the minimum of the validation error, which is not visible in the plot. The colored area shows the errors associated with the estimation of the validation error, while the error bar around the minimum and the dashed lines illustrate the `1-se rule', which points to $\tilde{\lambda} = 0.27$.}
\label{fig:cv1}
\end{figure}

\begin{figure}[t]
\centering
\includegraphics[width=0.45\textwidth]{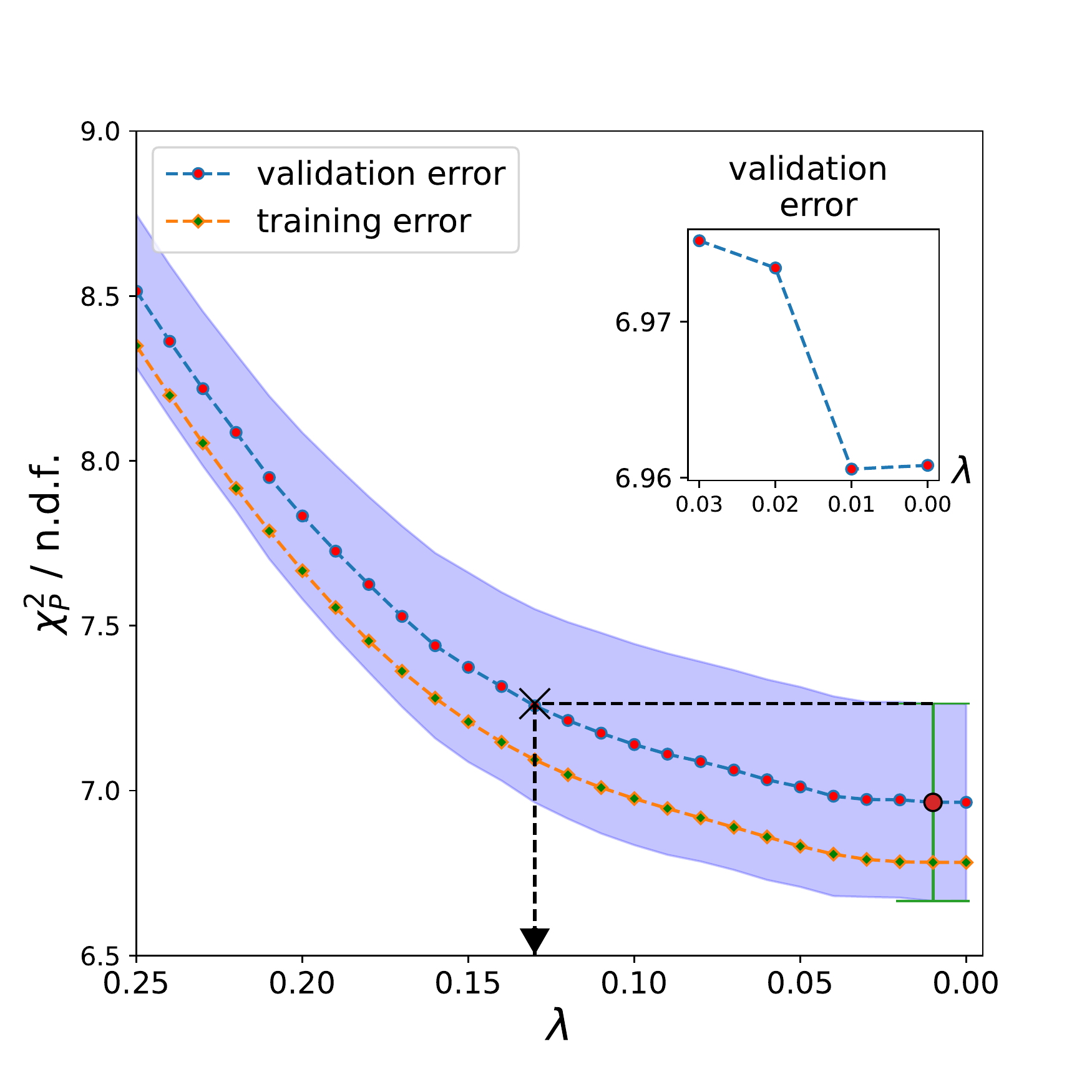}
\caption{Same as Fig.~\ref{fig:cv1}, but for the determination of the $\lambda$ value used in the BS2Mr fit ($\tilde{\lambda} = 0.13$).}
\label{fig:cv2}
\end{figure}

More precisely, after estimating the sample standard deviation of the $k$ validation errors 
\begin{equation}
  SD(\lambda) = \sqrt{Var(CV_{1}(\lambda),...,CV_{k}(\lambda))},
  \label{eq:SD}
\end{equation}
the standard error is computed as
\begin{equation}
  SE(\lambda) = SD(\lambda) / \sqrt{k}.
  \label{eq:SE}
\end{equation}
Thus, the new optimal $\tilde{\lambda}$, chosen according to the `1-se rule', is such that
\begin{equation}
\overline{CV}(\tilde{\lambda}) = \overline{CV}(\lambda^{*}) + SE(\lambda^{*}).
  \label{eq:lopt}
\end{equation}
Figs.~\ref{fig:cv1} and ~\ref{fig:cv2}, demonstrate the application of the `1-se rule' in two  different instances.

\begin{figure}
    \centering
    \includegraphics[width=0.5\textwidth]{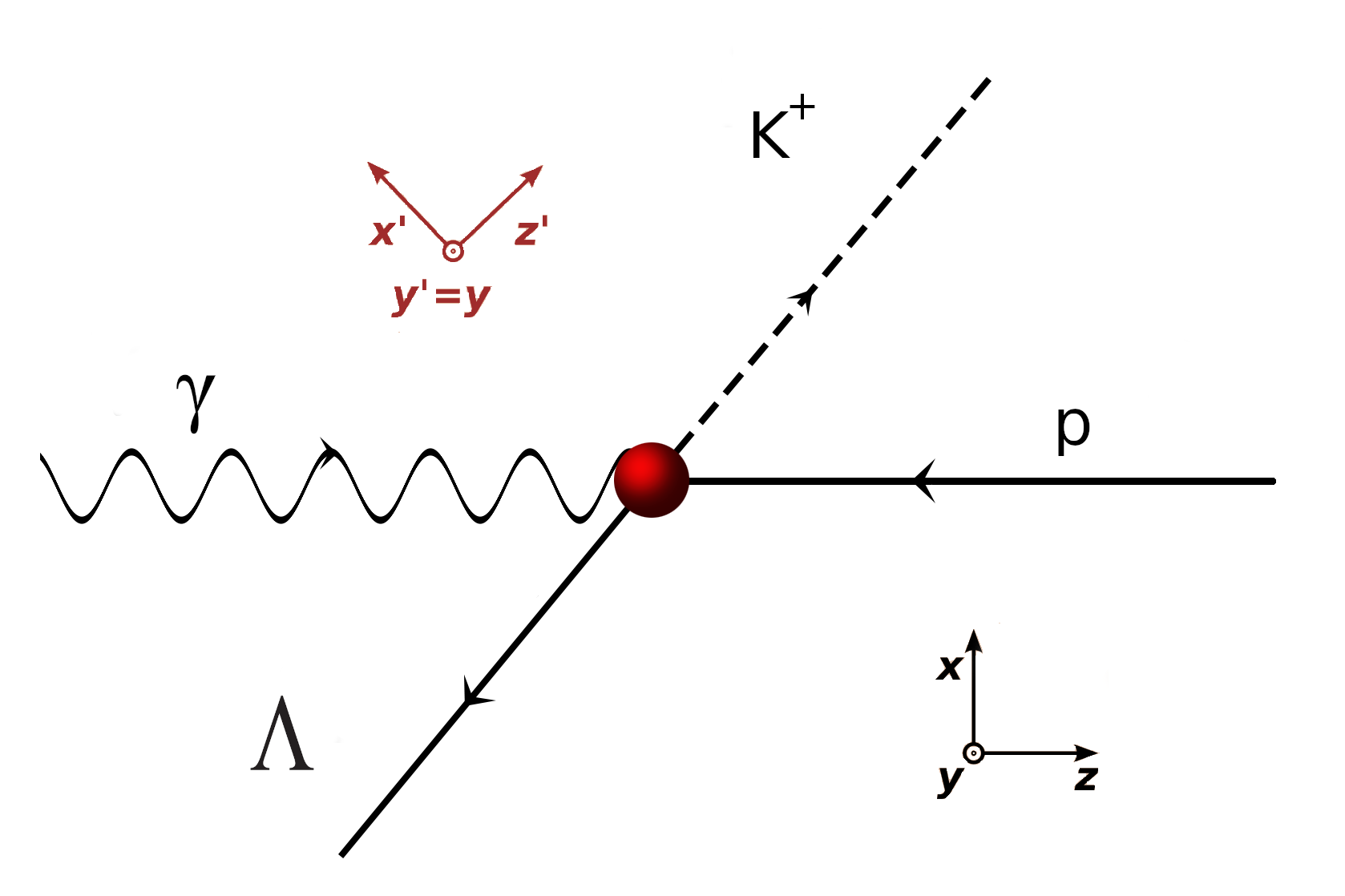}
    \caption{Kinematics of the $p(\gamma,K^+)\Lambda$ process together with the center-of-mass reference frames before (no primes) and after (primed) the reaction.}
    \label{fig:kinematics}
\end{figure}

\subsection{Experimental data}

In the fitting procedure, we have used experimental data on differential cross sections from the CLAS Collaboration~\cite{CLAS05,CLAS10}, which were limited in (center-of-mass) energy up to $W=2.355~\text{GeV}$, differential cross sections from the LEPS Collaboration~\cite{LEPS06} and we also used the differential-cross-section data collected in Ref.~\cite{AS}. Moreover, we used the hyperon polarization $P$ data from the CLAS Collaboration~\cite{CLAS10}, which were limited to $W = 2.225\,\text{GeV}$. These data, which we used previously in order to reach the BS1 and BS2 models, were replenished with several data sets from CLAS on photon beam polarization asymmetry $\Sigma$, target polarization asymmetry $T$ and beam-recoil polarization observables $O_{x^\prime}$ and $O_{z^\prime}$~\cite{Paterson}. These data span the energy range from $W = 1.71\,\text{GeV}$ to 2.19~GeV. In total, we have used 4640 datapoints to fit the free parameters of our model.

\begin{figure}[t]
\centering
\includegraphics[width=0.45\textwidth]{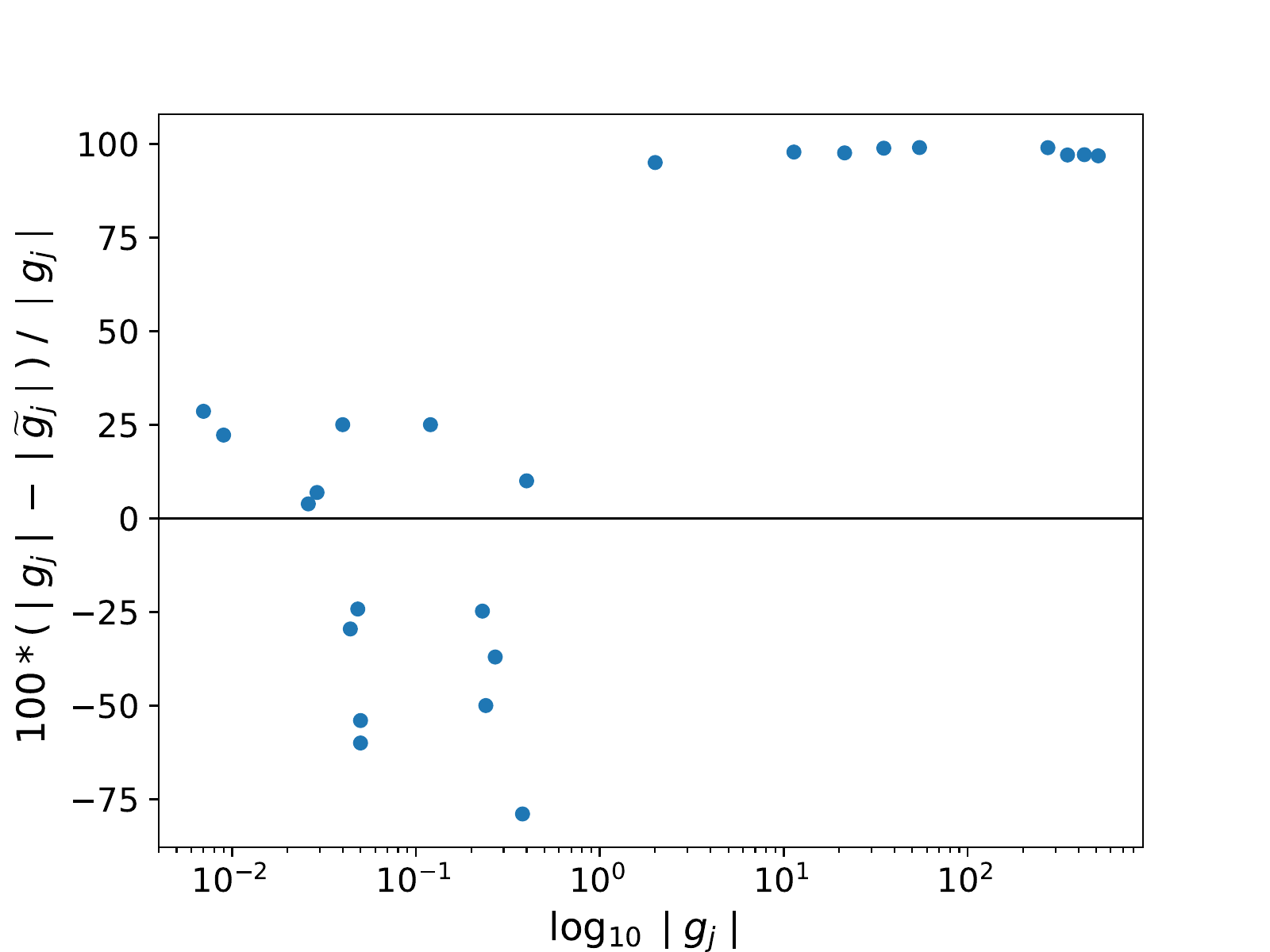}
\caption{Relative percentage reduction in absolute values of the resonance couplings as a function of the logarithm of their magnitudes, as a result of regularization. The $g_j$ values are the couplings that result from the unregularized BS2Mr0 fitting, while $\tilde{g}_j$ are the corresponding values after performing Ridge regularization in BS2Mr.}
\label{fig:cv3b}
\end{figure}

Single polarization asymmetries have the form
\begin{equation}
\frac{d\sigma^{+} - d\sigma^{-}}{d\sigma^{+} + d\sigma^{-}},
\label{eq:single}
\end{equation}
where $d\sigma$ denotes the differential cross section and the superscript  $(+\,\, \textrm{or}\,\, -)$ the polarization state (parallel or antiparallel) with respect to the corresponding quantization axis. So, the target polarization asymmetry ($T$) refers to the spin projection of the nucleon on the $y$-axis, while the recoil polarization ($P$) refers to the polarization of the hyperon on the $y^\prime$-axis.

The photon beam asymmetry 
\begin{equation}
\Sigma = \frac{d\sigma^{\bot} - d\sigma^{\parallel}}{d\sigma^{\textrm{unpol}}},
\label{eq:Sigma}
\end{equation}
refers to linearly polarized photons along the $x$-axis ($\bot$) and the $y$-axis ($\parallel$).

For the double polarization asymmetries, which are defined as
\begin{equation}
\frac{d\sigma^{(++)}+d\sigma^{(--)}-d\sigma^{(+-)}-d\sigma^{(-+)}}{d\sigma^{(++)}+d\sigma^{(--)}+d\sigma^{(+-)}+d\sigma^{(-+)}}
\label{eq:double}
\end{equation}
the polarizations of two particles are taken into account, which appear as superscripts in the $d\sigma$'s. In the case of beam-recoil asymmetries ($O_{x'}, O_{z'}$) that we examine in this work, the photons are linearly polarized along the direction that bisects the $x$- and $y$-axes, while the hyperon polarizations are along the $x^\prime$- and $z^\prime$-axes. The beam-recoil asymmetries ($C_x, C_z$) correspond to circularly polarized photons. The kinematics of the process together with the reference frames are shown in the Fig.~\ref{fig:kinematics}. A more rigorous treatment of polarization observables can be found in~Ref.\cite{Barker}.

\section{Results and discussion}
\label{sec:results}

The two fittings that we conducted using Ridge regularization are denoted BS2r and BS2Mr and they contain different sets of resonances (see discussion below). In both cases, we used 3-fold cross-validation in order to extract the optimal value for the regularization parameter $\lambda$ and the results are demonstrated in Figs.~\ref{fig:cv1} and \ref{fig:cv2}. As described in Subsec.~\ref{sub:cv}, the points depicting the training and validation errors represent the averages of the 3 cross validation runs, for each value of $\lambda$, while the optimum $\tilde{\lambda}$ is obtained by the `1-se rule'.



As can be seen in Fig.~\ref{fig:cv1}, in the case of BS2r, this process leads to a value $\tilde{\lambda} = 0.27$, while in the case of BS2Mr (Fig.~\ref{fig:cv2}) $\tilde{\lambda} = 0.13$. Using these $\lambda$ values in the penalty term of Eq.~\ref{eq:Pen2} and refitting the whole data set with the error function of Eq.~\ref{eq:chi^2_T} gives the BS2r and BS2Mr results, respectively. The corresponding versions of these two models with no regularization ($\lambda = 0$) are called BS2r0 and BS2Mr0.

In the BS1 and BS2 models of Ref.~\cite{BS2016} the couplings of spin-1/2 hyperon resonances acquire values around or above 10. With the help of Ridge regression, we were able to shrink most of them by an order of magnitude. 
The couplings of the spin-3/2 hyperon resonance $\Sigma(1940)$ are decreased (in absolute values) as well; they change from $g_1 = -0.86$ and $g_2 = 0.18$ to $g_1= -0.69$ and $g_2= 0.079$, respectively. In this case the difference is not so large as in the case of spin-1/2 hyperon couplings. In general, it can be said that the larger couplings get penalized more. The results of this fitting are summarized in Table~(\ref{tab:BS2r}). Compared to its unregularized version (BS2r0), the $\chi^2$ error in BS2r is increased by $\sim$ 9 \%, while the hyperon couplings are reduced on average by 70 \%.

\begin{widetext}

\begin{table}[h]
\caption{Properties of considered resonances in the BS2r and BS2Mr fits. We show masses, widths, and couplings of the included resonances. Labels of the resonances are the same as in \cite{BS2016}.}
\begin{center}
\begin{tabularx}{\textwidth} {llrrrrrr}
\hline \hline
Label  & Resonance           & Mass [MeV] & Width [MeV] & $g_1$ (BS2r)  & $g_2$ (BS2r) & $g_1$ (BS2Mr)  & $g_2$ (BS2Mr)\\
\hline                                                                    
K*& $K^*(892)$             & 891.7      &  50.8       & $-0.163 \pm 0.002 $ & $ 0.045 \pm 0.003$    & $-0.098 \pm 0.001 $ & $ -0.249 \pm 0.005$ \\ 
K1& $K_1(1272)$            & 1272       &  90         & $ 0.286 \pm 0.004 $ & $-1.042 \pm 0.010$    & $ 0.521 \pm 0.006 $ & $ -0.550 \pm 0.008$ \\
N3& $N(1535)\;1/2^-$       & 1530       & 150         & $ 0.002 \pm 0.005 $ & $ -     $    & $ 0.684 \pm 0.004 $ & $ -     $ \\
N4& $N(1650)\;1/2^-$       & 1650       & 125         & $-0.078 \pm 0.002 $ & $ -     $    & $-0.287 \pm 0.001 $ & $ -     $ \\
P5& $N(1860)\;5/2^+$       & 1860       & 270         & $-0.014 \pm 0.001 $ & $ 0.006 \pm 0.001$    & $ 0.057 \pm 0.001 $ & $ -0.060 \pm 0.001$ \\
N7& $N(1720)\;3/2^+$       & 1720       & 250         & $ 0.145 \pm 0.002 $ & $ 0.0001 \pm 0.0004$    & $ 0.089 \pm 0.002 $ & $ 0.027 \pm 0.0004$ \\
P4& $N(1875)\;3/2^-$       & 1875       & 265         & $ 0.116 \pm 0.002 $ & $ 0.112 \pm 0.001$    & $ 0.360 \pm 0.002 $ & $ 0.367 \pm 0.002$ \\
P2& $N(1900)\;3/2^+$       & 1920       & 200         & $-0.028 \pm 0.001 $ & $-0.005 \pm 0.0003$    & $ 0.025 \pm 0.001 $ & $ -0.031 \pm 0.0003$ \\
P3& $N(2050)\;5/2^+$       & 2050       & 220         & $-0.012 \pm 0.0002$ & $ 0.012 \pm 0.0002$    & $-0.007 \pm 0.0001 $ & $ 0.005 \pm 0.0001$ \\
N9& $N(1685)\;5/2^+$       & 1685       & 130         & $ 0.046 \pm 0.001 $ & $-0.038 \pm 0.001$    & $-0.081 \pm 0.001 $ & $ 0.077 \pm 0.001$ \\
N6& $N(1710)\;1/2^+$       & 1710       & 140         & $-0.141 \pm 0.004 $ & $ -     $    & $-0.358 \pm 0.004 $ & $ -     $ \\
L1& $\Lambda(1405)\;1/2^-$ & 1405       &  51         & $ 2.624 \pm 0.078 $ & $ -     $    & $ 12.52 \pm 0.12 $ & $ -     $ \\
S1& $\Sigma(1660)\;1/2^+$  & 1660       & 100         & $-5.925 \pm 0.126 $ & $ -     $    & $-10.47 \pm 0.50 $ & $ -     $ \\
L2& $\Lambda(1405)\;1/2^+$ & 1600       & 150         &                     &              & $ -16.34 \pm 0.44 $ & $ -     $ \\
L4& $\Lambda(1800)\;1/2^-$ & 1800       & 300         & $-1.409 \pm 0.161 $ & $ -     $    &                   &            \\
S4& $\Sigma(1940)\;3/2^-$  & 1940       & 220         & $-0.685 \pm 0.022 $ & $ 0.079 \pm 0.005$    & $-3.08 \pm 0.04 $ & $ 0.408 \pm 0.010$ \\
L5& $\Lambda(1810)\;1/2^+$ & 1810       & 150         &                     &              & $-2.83 \pm 0.76 $ & $ -     $ \\

\hline \hline
\end{tabularx}

\end{center}
\label{tab:BS2r}
\end{table}

\end{widetext}

Motivated by Ref.~\cite{Mart-are} we found that when we replace $\Lambda(1800)1/2^-$ (L4) that is used in the BS2 model with $\Lambda(1600)1/2^+$ (L2) and $\Lambda(1810)3/2^+$ (L5), while keeping the rest of the resonance set from the BS2 model, we describe the polarization observables in some kinematic regions much better and reproduce the data very aptly. This set of resonances is used in the BS2Mr and BS2Mr0 fits. In the fit without regularization (BS2Mr0) the hyperon couplings acquire extremely large values, which decrease substantially after the introduction of regularization in BS2Mr. Fig.~\ref{fig:cv3b} shows the effect of regularization in the absolute values of the couplings. One sees that couplings with magnitudes greater than 1 (which correspond to the kaons and the hyperons) get decreased by over 90 \%. The rest of the couplings (corresponding to the nucleons) undergo more moderate reductions, or even increase in some cases, but still remain within physically acceptable limits.


In the MINUIT fitting procedure, the variable parameters with limits undergo a transformation which is non-linear and thus it is recommended to avoid constraining the parameters if it is not needed~\cite{Minuit-writeup}. It is the Ridge regression which can help us preventing this issue as the penalty function imposed on the parameters restricts their values instead of the imposed limits. If we let the hyperon couplings vary freely, they tend to acquire large values leading to large contributions of corresponding amplitudes. The tool to suppress these contributions in our model is the hadronic form factor, particularly its cut-off parameter $\Lambda$; the smaller its value the stronger the suppression. But in the fits with large hyperon couplings we have unphysically small cut-off parameter. Before opting for Ridge regression technique, the only way to deal with this problem was imposing limits on the cut-off parameter. However, once we penalize the hyperon couplings, we end up with a model, which has not only decreased hyperon couplings but also larger and thus more physically acceptable value of the cut-off parameter. In general, one can therefore say that the use of Ridge regression leads to more physically acceptable values of the fitted parameters.

In Fig.~\ref{fig:Sg-all}, we show the photon-beam asymmetry data and the corresponding predictions of BS2r, BS2Mr, BS2r0, and BS2Mr0 fits. The BS2r fit captures the shape of the data reasonably well from the threshold up to around 1.9~GeV. Above 2~GeV the BS2r fit can still capture the sign and magnitude of the data above $\cos\theta_K^{c.m.} = -0.5$, whereas it fails to describe data below $\cos\theta_K^{c.m.} = -0.5$ -- while the data are around 0.5, the model gives photon-beam asymmetry which is of the same magnitude but opposite sign.  Above $W = 2.1\,\text{GeV}$ the BS2r fit cannot reproduce the peak around central angles. The BS2r0 version behaves in a similar way, but it underestimates the data at energies above 2.0~GeV and then above $W=2.1\,\text{GeV}$, where there is a peak in the data, it gives rather a dip. Up to 2~GeV, the BS2Mr fit works in a similar fashion to the BS2r fit, above 2~GeV it gives beam asymmetries which are flat. Interestingly, the BS2r fit is the only fit that can reproduce the shape in the beam-asymmetry data around $W=2\,\text{GeV}$.

The target polarization asymmetry data and their description by our fits are shown in Fig.~\ref{fig:T-all}. At energies below 2~GeV, the BS2r and BS2r0 fits behave in a similar manner but above 2~GeV they can capture the shapes in the data at backward angles giving rise to a broad peak. The two other fits, BS2Mr and BS2Mr0, give similar description of target polarization asymmetry data as the BS2r and BS2r0.

The beam-recoil asymmetry data, $O_{x^\prime}$, are compared with our fits in Fig.~\ref{fig:Ox-all}. The BS2r and BS2r0 fits captures the data above $W=1.9\,\text{GeV}$ and in the forward kaon angles quite well. A slightly different description of data can be seen from the BS2Mr and BS2Mr0 fits. From $W=1.9\,\text{Gev}$ they can capture the data at forward angles and beyond 2~GeV they give overall good description of all available $O_{x^\prime}$ data.

Another set of beam-recoil asymmetry data, $O_{z^\prime}$, and results of our fits are plotted in Fig.~\ref{fig:Oz-all}. The BS2r and BS2r0 are a bit more successful in describing $O_{z^\prime}$ data than they are in describing the $O_{x^\prime}$ data as they more or less agree with the data for $W>1.8\,\text{GeV}$ at forward kaon angles. Reliable description of $O_{z^\prime}$ data at backward angles beyond 2~GeV is given by the BS2Mr and BS2Mr0 fits.

In our analysis, we did not fit the double-polarization data $C_x$ and $C_z$ measured by the CLAS Collaboration~\cite{CLAS2007}. Instead, we used these data for testing the predictive power of our results. In Figures~\ref{fig:Cx} and \ref{fig:Cz} we collect experimental data and model predictions on $C_x$ and $C_z$, respectively.  In general, we see better agreement with the data of both the BS2Mr and BS2Mr0 predictions than the BS2r and BS2r0 predictions, especially in the case of $C_z$, Fig.~\ref{fig:Cz}. This may prove the superiority of the BS2M set of hyperon resonances for description of double-polarization data.

Among the hyperon resonances, the most important part is played by the $\Lambda(1405)$ which changes the model predictions substantially once omitted. Without this resonance, the BS2r fit beam asymmetries drop almost to zero and no longer give the shape that can be seen in the data. The agreement with the target-asymmetry and $O_{x^\prime}$ data is also severed as the fit does not catch the shapes in the data. Only in the case of $O_{z^\prime}$ data the correspondence between fit and experiment does not change much. The other hyperon resonances which we include in the BS2r fit, L4, S1, and S4 apparently do not play decisive roles in the data description since when we omit them, we notice only very slight changes in how the fit agrees with data.

\begin{widetext}

\begin{figure}
    \centering
    \includegraphics[width=0.8\textwidth]{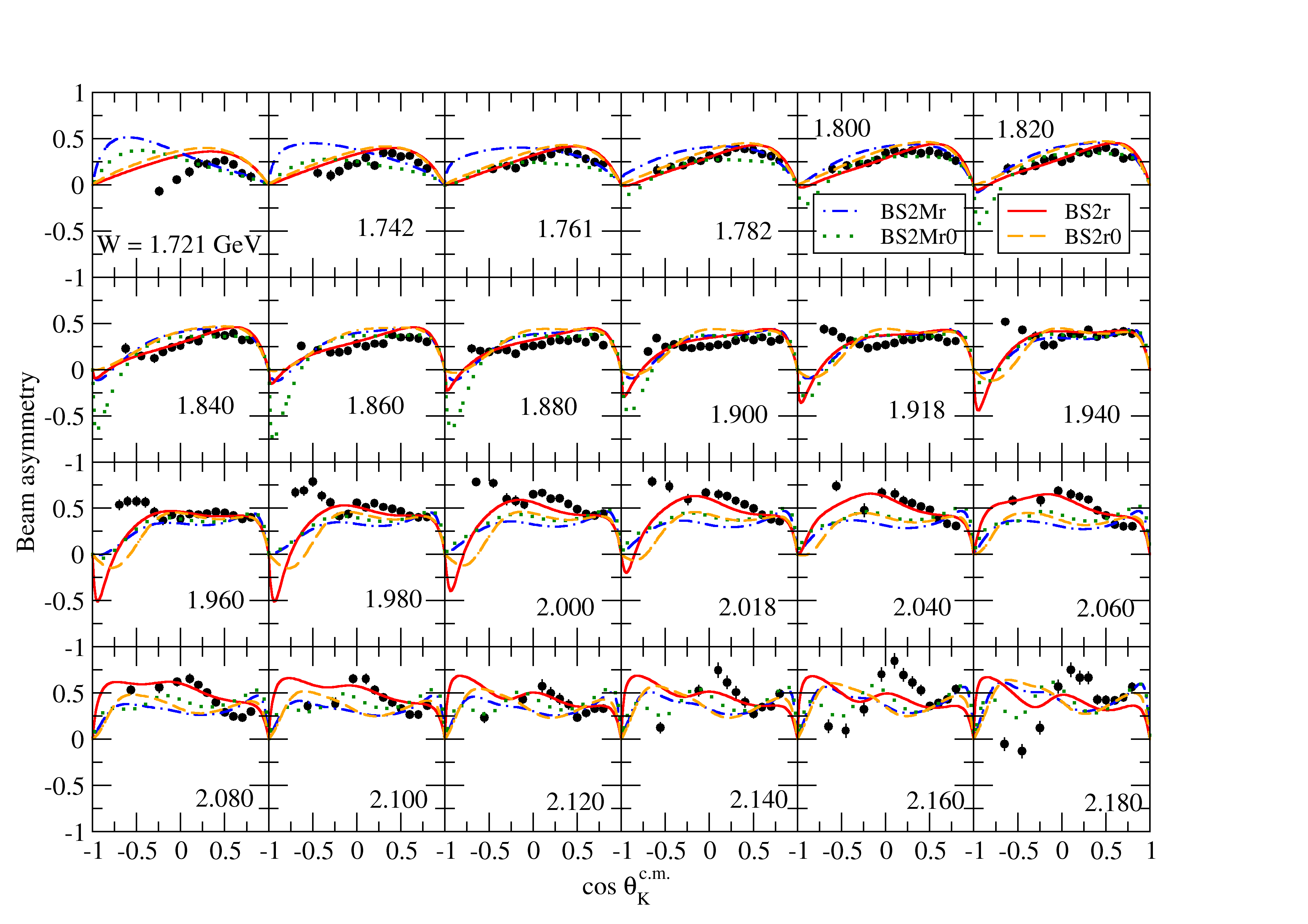}
    \caption{Photon-beam asymmetry data measured by CLAS~\cite{Paterson} in comparison with the BS2r (solid line), BS2r0 (dashed line), BS2Mr (dash-dotted line), and BS2Mr0 (dotted line) fits.}
    \label{fig:Sg-all}
\end{figure}

\begin{figure}
    \centering
    \includegraphics[width=0.8\textwidth]{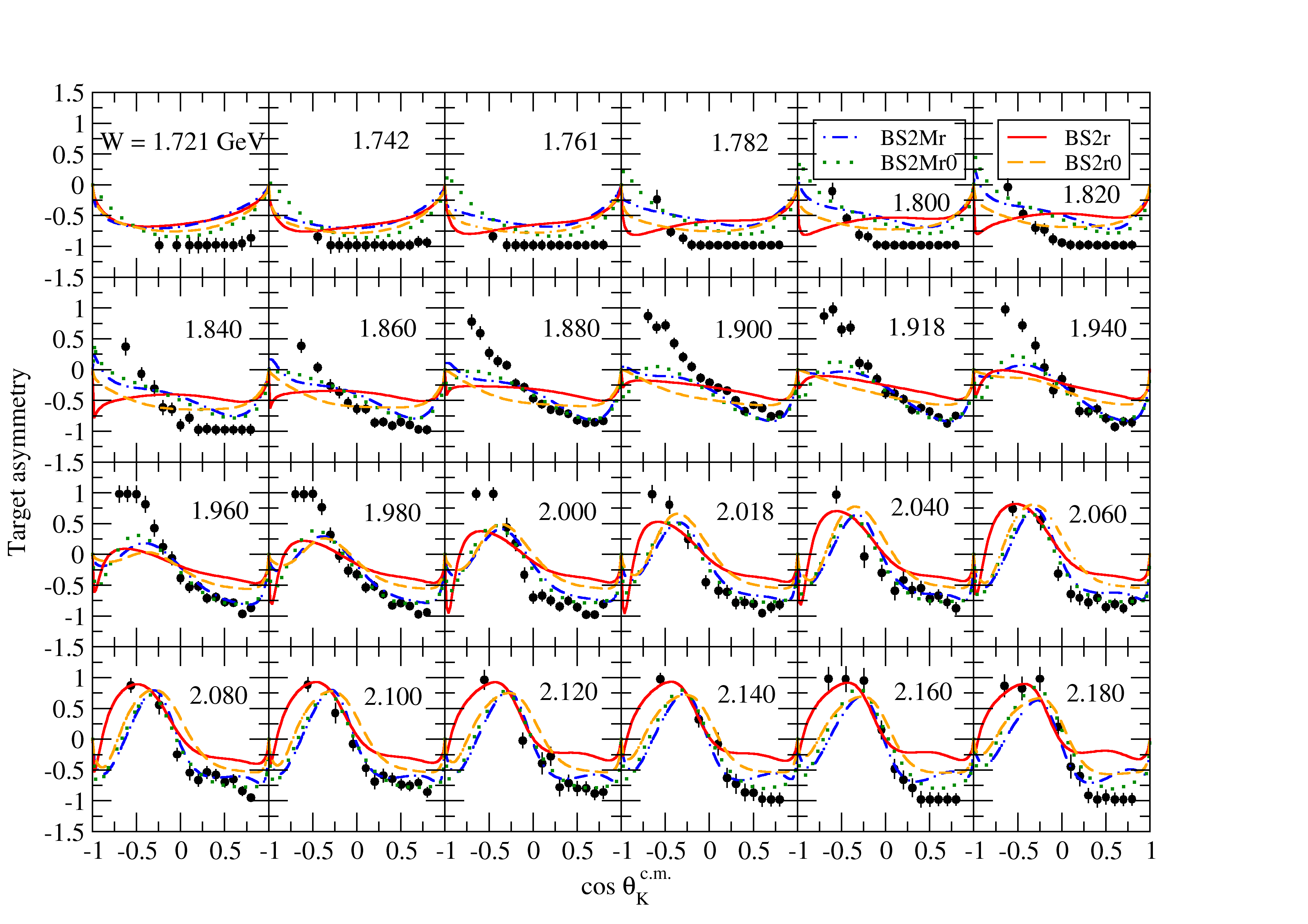}
    \caption{Target asymmetry data measured by CLAS~\cite{Paterson} in comparison with the BS2r (solid line), BS2r0 (dashed line), BS2Mr (dash-dotted line), and BS2Mr0 (dotted line) fits.}
    \label{fig:T-all}
\end{figure}

\begin{figure}
    \centering
    \includegraphics[width=0.8\textwidth]{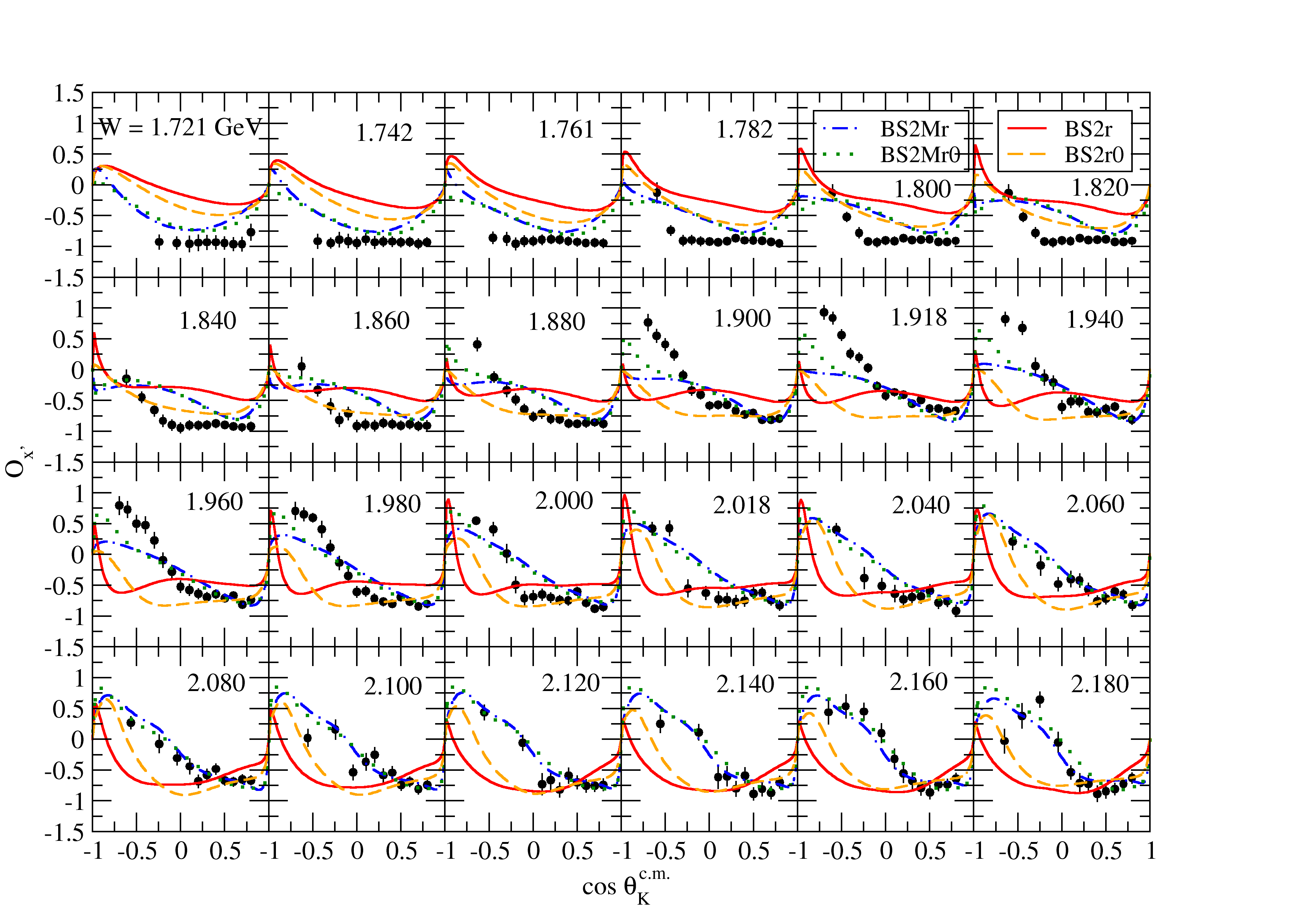}
    \caption{Double-polarization asymmetry, $O_{x^\prime}$, data measured by CLAS~\cite{Paterson} in comparison with the BS2r (solid line), BS2r0 (dashed line), BS2Mr (dash-dotted line), and BS2Mr0 (dotted line) fits.}
    \label{fig:Ox-all}
\end{figure}

\begin{figure}
    \centering
    \includegraphics[width=0.8\textwidth]{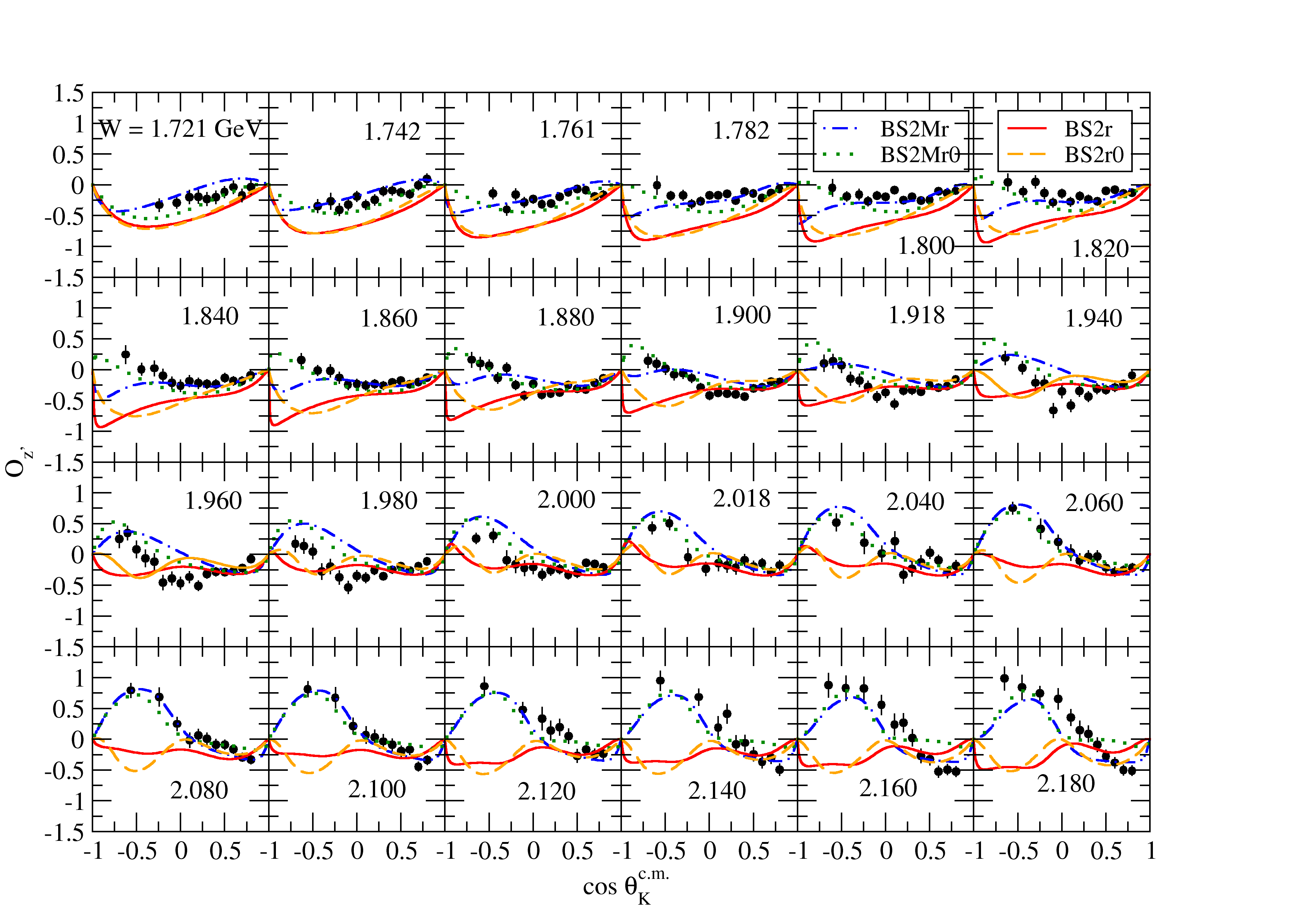}
    \caption{Double-polarization asymmetry, $O_{x^\prime}$, data measured by CLAS~\cite{Paterson} in comparison with the BS2r (solid line), BS2r0 (dashed line), BS2Mr (dash-dotted line), and BS2Mr0 (dotted line) fits.}
    \label{fig:Oz-all}
\end{figure}

\begin{figure}
    \centering
    \includegraphics[width=0.8\textwidth]{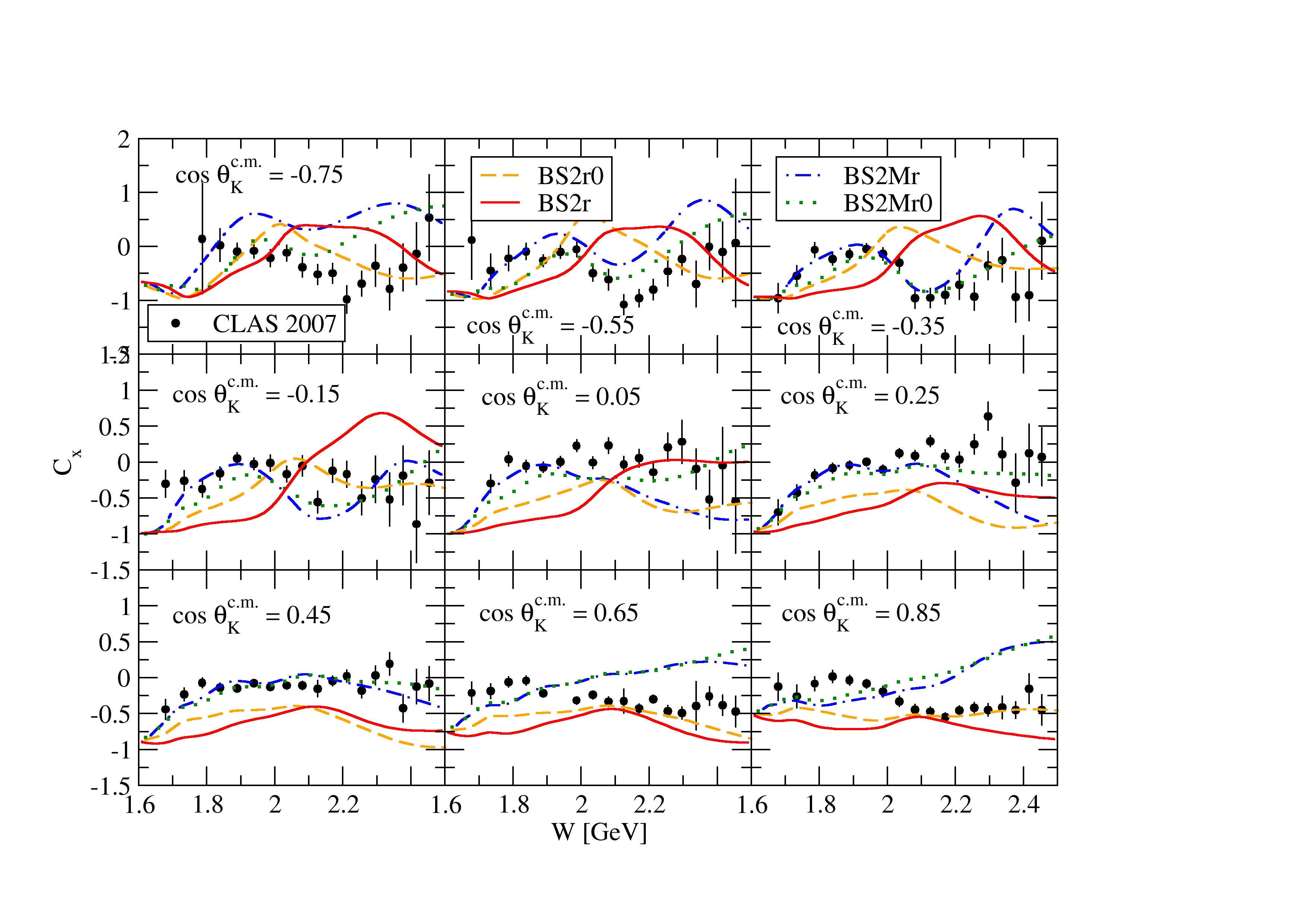}
    \caption{Double-polarization asymmetry, $C_{x}$, data measured by CLAS~\cite{CLAS2007} in comparison with predictions of the BS2r (solid line), BS2r0 (dashed line), BS2Mr (dash-dotted line), and BS2Mr0 (dotted line) fits.}
    \label{fig:Cx}
\end{figure}

\begin{figure}
    \centering
    \includegraphics[width=0.8\textwidth]{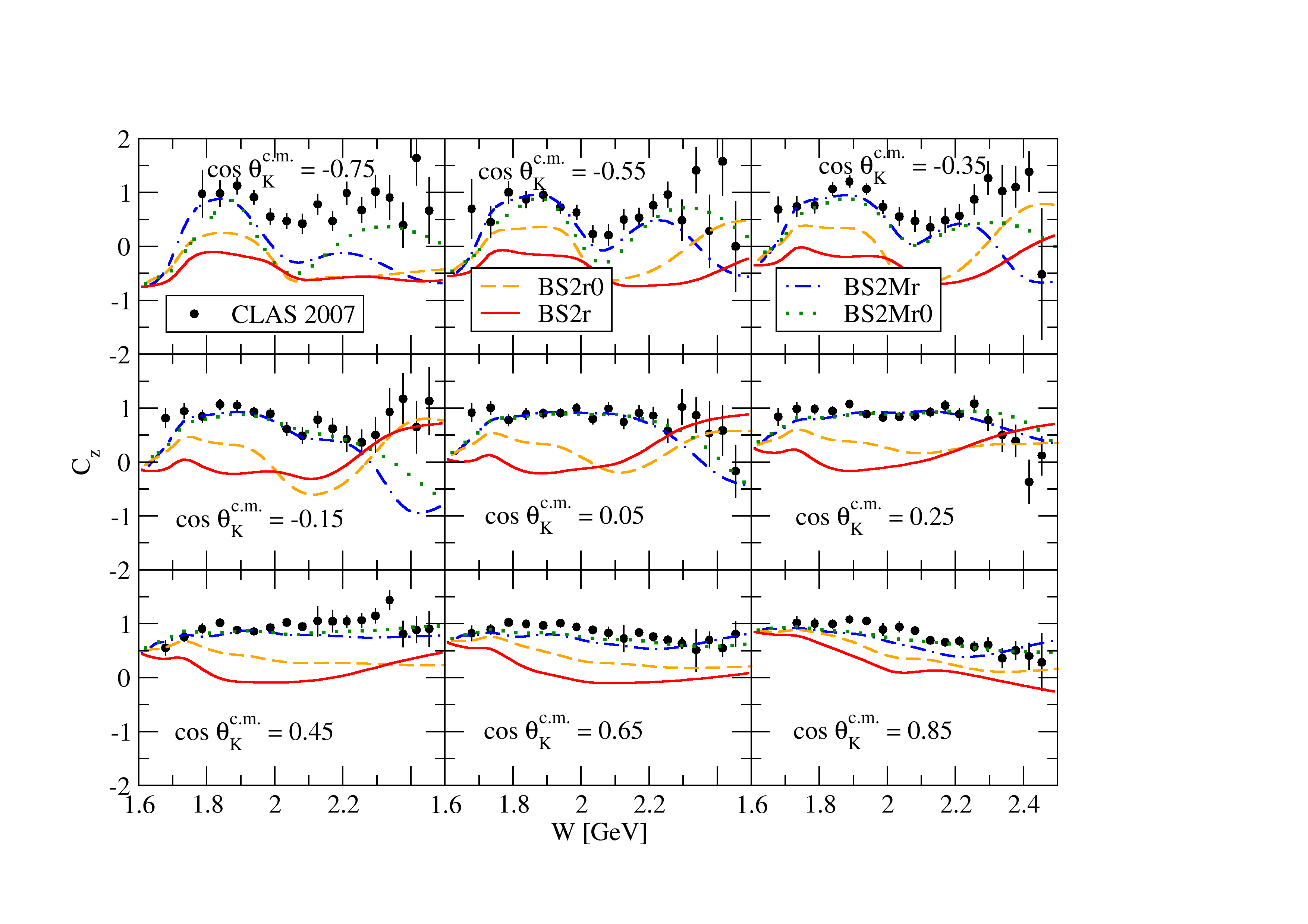}
    \caption{Double-polarization asymmetry, $C_{z}$, data measured by CLAS~\cite{CLAS2007} in comparison with predictions of the BS2r (solid line), BS2r0 (dashed line), BS2Mr (dash-dotted line), and BS2Mr0 (dotted line) fits.}
    \label{fig:Cz}
\end{figure}

\end{widetext}





\section{Conclusion}
\label{sec:conclusion}

In this paper, we used the isobar model and studied the photoproduction of $K^+\Lambda$ off a proton target in the resonance region. The amplitude in the isobar model is constructed in the tree level with help of effective meson-baryon Lagrangians. In the corresponding Feynman diagrams we assume exchanges of particles in their ground as well as excited states. 


The most important part of this analysis lies in the enhancement of the fitting method. In machine learning, the standard method to avoid overfitting the data is regularization. In this method, a constraint is added to the $\chi^2$ which, depending on its shape, either forces parameters to take zero values, which can be used as a suitable model selection tool (LASSO), or shrinks the parameters without annihilating them -- a method which is known under the name Ridge regression. In this work we used the Ridge regression and applied this technique to our previously published BS2 model. Our goal was not to find another model but rather to modify overly large couplings of hyperon resonances present in the BS2 model. With the help of Ridge, we could suppress their values by an order of magnitude to more physically acceptable values. 

Unlike LASSO, which serves as a tool for model selection, Ridge can be used within one's model of choice. Apart from the obvious benefit of preventing the couplings from obtaining large values, a regularized model is less prone to overfitting and therefore can generalize better to new data. We believe that these gains more than compensate for the increase in error that entails the use of regularization.

Subsequently, we studied the contributions of hyperon resonances and tried to disentangle the complicated structure of the background to the $K^+\Lambda$ photoproduction. This led us to replacing the $\Lambda(1800)1/2^-$ (L4) hyperon resonance with $\Lambda(1600)1/2^+$ (L2) and $\Lambda(1810)3/2^+$ (L5) hyperon resonances which resulted into a slightly better agreement with data in some kinematic regions. We identified the $\Lambda(1405)$ as the most important hyperon resonance for description of polarization observables.

Obtaining a more realistic description of background allows us to perform a more reliable analysis of the resonant part of the amplitude, particularly on the role and importance of the nucleon resonances.

In the near future, we would like to concentrate on the analysis of $\Sigma$ photoproduction channels. To this end, we plan to use both regularization methods mentioned here; LASSO for selecting the most appropriate set of resonances contributing to the process and Ridge, if needed, for naturally reducing the values of fitted couplings.

\section{Acknowledgements}
The authors would like to thank P.~Byd\v{z}ovsk\'{y} for careful reading of the manuscript and useful comments.



\begin{thebibliography}{10}

\bibitem{hyper}
P.~Byd\v{z}ovsk\'{y}, M. Sotona, T. Motoba, K. Itonaga, K. Ogawa, and O. Hashimoto, Nucl. Phys. A \textbf{881}, 199 (2012); T. Motoba, P. Byd\v{z}ovsk\'{y}, M. Sotona, and K. Itonaga, Prog. Theor. Phys. Suppl. \textbf{185}, 224 (2010); P. Byd\v{z}ovsk\'{y} and T. Mart, Phys. Rev. C \textbf{76}, 065202 (2007).

\bibitem{Kuo}
T.~K.~Kuo, Phys. Rev. \textbf{129}, 2264 (1963)

\bibitem{Thom}
H.~Thom, Phys. Rev. \textbf{151}, 1322 (1966).

\bibitem{ABW}
R.~A.~Adelseck, C.~Bennhold, L.~E.~Wright, Phys. Rev. C \textbf{32}, 1681 (1985).


\bibitem{Rosenthal}
A.~S.~Rosenthal \emph{et al.}, Ann. Phys. (N.Y.) \textbf{184}, 33 (1988).

\bibitem{AS}
R.~A.~Adelseck, B.~Saghai, Phys. Rev. C \textbf{42}, 108 (1990).

\bibitem{WJC}
R.~A.~Williams, Chueng-Ryong Ji, and S.~R.~Cotanch, Phys. Rev. C \textbf{46}, 1617 (1992).

\bibitem{SL}
J.~C.~David, C.~Fayard, G.-H.~Lamot, and B.~Saghai, Phys. Rev. C \textbf{53}, 2613 (1996).

\bibitem{Mizutani}
T.~Mizutani, C.~Fayard, G.-H.~Lamot, and B.~Saghai, Phys. Rev. C \textbf{58}, 75 (1998).

\bibitem{Mart-Bennhold}
T.~Mart, C.~Bennhold, Phys. Rev. C \textbf{61}, 012201(R) (1999).

\bibitem{Guidal}
M.~Guidal, J.-M.~Laget, M.~Vanderhaeghen, Nucl. Phys. A \textbf{627}, 645-678 (1997).

\bibitem{GentIM}
S.~Janssen, J.~Ryckebusch, D.~Debruyne, and T.~Van~Cauteren, Phys. Rev. C \textbf{65}, 015201 (2001).

\bibitem{Janssen-2003}
S.~Janssen, D.~G.~Ireland, J.~Ryckebusch, Phys. Lett. B \textbf{562}, 51 (2003).

\bibitem{Janssen-2001EPJ}
S.~Janssen, J.~Ryckebusch, W.~Van Nespen, D.~Debruyne, and T.~Van Cauteren, Eur. Phys. J. A \textbf{11}, 105 (2001).

\bibitem{Mart-are}
T.~Mart, and N.~Nurhadiansyah, Few-Body Syst. \textbf{54}, 1729–1739 (2013).

\bibitem{CLAS05}
R.~Bradford \emph{et al.}, Phys. Rev. C \textbf{73}, 035202 (2006); J.~W.~C. McNabb \emph{et al.} (CLAS Collaboration) \emph{ibid.} \textbf{69}, 042201(R) (2004).

\bibitem{Puente}
A.~de la Puente, O.~V.~Maxwell, B.~A.~Raue, Phys. Rev. C \textbf{80}, 065205 (2009).

\bibitem{Corthals}
T.~Corthals, J.~Ryckebusch, and T.~Van Cauteren, Phys. Rev. C \textbf{73}, 045207 (2006).

\bibitem{DeCruz}
L.~De Cruz, J.~Ryckebusch, T.~Vrancx, P.~Vancraeyveld, Phys. Rev. C \textbf{86}, 015212 (2012).

\bibitem{Shklyar}
V.~Shklyar, H.~Lenske, and U.~Mosel, Phys. Rev. C \textbf{72}, 015210 (2005).

\bibitem{Shyam}
R.~Shyam, O.~Scholten, and. H.~Lenske, Phys. Rev. C \textbf{81}, 015204 (2010).

\bibitem{Julia}
B.~Julia-Diaz, B.~Saghai, T.-S.~H.~Lee, and F.~Tabakin, Phys. Rev. C \textbf{73}, 055204 (2006).

\bibitem{Borasoy}
B.~Borasoy, P.~C.~Bruns, U.-G.~Meissner, and R.~Nissler, Eur. Phys. J. A \textbf{34}, 161 (2007).

\bibitem{BG1}
A.~V.~Sarantsev, V.~A.~Nikonov, A.~V.~Anisovich, E.~Klempt, and U.~Thoma, Eur. Phys. J. A \textbf{25}, 441 (2005).

\bibitem{BG2}
A.~V.~Anisovich, V.~Kleber,  E.~Klempt, V.~A.~Nikonov, A.~.V.~Sarantsev, and U.~Thoma, Eur. Phys. J. A \textbf{34}, 243 (2007).

\bibitem{BG2012}
A.~V.~Anisovich, R.~Beck,  E.~Klempt, V.~A.~Nikonov, A.~V.~Sarantsev, and U.~Thoma, Eur. Phys. J. A \textbf{48}, 15 (2012).

\bibitem{BG3}
A.~V.~Anisovich, \emph{et al.}, Eur. Phys. J. A \textbf{53}, 242 (2017).

\bibitem{Roenchen}
D.~Roenchen, M.~Doering, and U.-G.~Meissner, Eur. Phys. J. A \textbf{54}, 110 (2018).

\bibitem{Kamano}
H.~Kamano, S.~X.~Nakamura, T.-S.~H.~Lee, and T.~Sato, Phys. Rev. C \textbf{88}, 035209 (2013).

\bibitem{JPARC}
H.~Noumi \emph{et al.}: ”Spectroscopic study of hyperon resonances below $K\bar{N}$ threshold via the $(K^{-},n)$ reaction on deuteron", proposal for the J-PARC 50 GeV Proton Synchrotron, submitted to the 8th PAC meeting, July 2009, available at http://j-parc.jp/researcher/Hadron/en/pac\_0907/pdf/Noumi.pdf

\bibitem{BS2016}
D.~Skoupil, P. Byd\v{z}ovsk\'{y}, Phys. Rev. C \textbf{93}, 025204 (2016).

\bibitem{Paterson}
C.~A.~Paterson \emph{et al.}, (CLAS Collaboration), Phys. Rev. C \textbf{93}, 065201 (2016).

\bibitem{Pascalutsa}
V.~Pascalutsa, Phys. Rev. D \textbf{58}, 096002 (1998).

\bibitem{Vrancx}
T.~Vrancx, L.~De Cruz, J.~Ryckebusch, and P.~Vancraeyveld, Phys. Rev. C \textbf{84}, 045201 (2011).

\bibitem{BS2018}
D.~Skoupil, P. Byd\v{z}ovsk\'{y}, Phys. Rev. C \textbf{97}, 025202 (2018).

\bibitem{Bishop}
C. M. Bishop, {\it Pattern Recognition and Machine Learning}, Springer (2006).

\bibitem{Hastie}
T.~Hastie, R.~Tibshirani and J.~Friedman, \textit{The Elements of Statistical Learning: Data Mining, Inference, and Prediction}, 2$^{nd}$ ed., Springer Series in Statistics, New York (2009).

\bibitem{Guegan}
B.~Guegan, J.~Hardin, J.~Stevens and M.~Williams, JINST \textbf{10}, P09002 (2015). 

\bibitem{Landay1}
J.~Landay, M.~D\"{o}ring, C.~Fern\'{a}ndez-Ram\'{i}rez, B.~Hu and R.~Molina, Phys. Rev. C \textbf{95}, 015203 (2017). 

\bibitem{Landay2}
J.~Landay, M.~Mai, M.~D\"{o}ring, H.~Haberzettl and K.~Nakayama, Phys. Rev. D \textbf{99}, 016001 (2019).

\bibitem{Bydzovsky}
P.~Byd\v{z}ovsk\'{y}, A.~Ciepl\'{y}, D.~Petrellis, D.~Skoupil and N.~Zachariou, Phys. Rev. C \textbf{104}, 065202 (2021).

\bibitem{DeCruzPhD}
L. De Cruz, Ph.D. thesis, Ghent University, 2011.

\bibitem{CLAS10}
M.~E.~McCracken \emph{et al.}, Phys. Rev. C \textbf{81}, 025201 (2010).

\bibitem{LEPS06}
M.~Sumihama \emph{et al.}, Phys. Rev. C \textbf{73}, 035214 (2006).

\bibitem{Barker}
I.~S.~Barker, A.~Donnachie, J.~K.~Storrow, Nucl. Phys. B\textbf{95} (1975) 347-356.

\bibitem{Minuit-writeup}
F.~James, MINUIT: Function Minimization and Error Analysis, CERN Program Library Long Writeup D506 available at: https://cdsweb.cern.ch/record/2296388/files/minuit.pdf.

\bibitem{CLAS2007}
R.~K.~Bradford \emph{et al.}, Phys. Rev. C 75, 035205 (2007).



\end{thebibliography}
\end{document}